# CORRELATION OF ASPECTS OF RECENT DRAWINGS MADE BY PHYSICISTS

Harm H. Hollestelle, Amsterdam, The Netherlands

ABSTRACT

In this study recent drawings made by physics students are investigated. These drawings are part of reports made during experimenting. Aspects of drawings are chosen for investigation that are linked to basic physical notions like plasticity and spatiality. Also the aspects are required to link to the motor recognition experience during drawing, of the one who made the drawing. In this way the aim is to clarify the individual notions of space and time that are basic to performing the experiments. These are called the attitude with which the experiment is performed. Some aspects of these drawings tend to be correlated in several trends. Polanyi's theory of consciousness is generalised to explain why this correlation of aspects occurs. The correlation trends and attitudes can be related to each other. The results are presented in a schematic way to facilitate interpretation of the concepts introduced and of the correlation results.

1. INTRODUCTION

In the past already much research has been done on drawings by physicists. Several authors have described the use of drawings in science both in general and in specific cases, in the present and in the past, such as Tufte [1], Lynch and Woolgar [2], and Baigrie [3]. They collected evidence about the importance of drawings made by physicists (and scientists in general) within pictorial, sociological and historical descriptions respectively.
Students drawings are used and investigated in other recent studies for several different reasons. They are used for psychological evaluation at the end of a years curriculum [4]. They are used to discuss and to improve on visualization of data results of experiments in general in science classes [5]. And they are used to check the improvement of understanding of sections of science theory [6,7].

The approach of Pittman [7, and references therein] is of special interest for this research. It is mentioned there: 'For new knowledge to be understood and remembered it must be meaningful to the learner. Meaningfulness depends on the learner's success in finding or creating connections between new information and pre-existing knowledge. One way by which these connections are made is through the use of analogies.' This is the same view as advocated by Read [8] who maintains that drawing is especially useful for gaining understanding because the use of analogies and the recognition of patterns and combinations in general is done always just in the same way as in drawing.

Drawings are, in the above approach, introduced as providing analogies for theories in science education. Analogies play a key-role in the historical development of scientific knowledge [9,10]. Also, concepts in the exact sciences can be stimulated by discussion from other branches of science [11]. In the research by Pittman analogies are allowed to be student-generated analogies containing 'outside of school' content like for instance building a house as an analogy for protein synthesis. Of course then in this case some steps of protein synthesis must find their way somehow into the drawing to make the analogy meaningful.

This research differs from the above mentioned ones because it attempts to describe drawings by physicists quantitatively and explicit properties, aspects of the drawings, are defined and measured and counted. These aspects are chosen in such a way that they relate to motor experience and time behaviour during drawing in order to investigate the spatial and temporal characteristics of making the drawing and of the drawing itself. By counting elementary parts of a drawing, parameters connected to the aspects are attributed to it.
Another difference is that visualization of results of physics experiments is not considered here, only drawings in reports of physics experiments that represent the experimental set-up or the theoretical situation of the experiment. Analogies will be important in this study, not to relate to content, but when trends of correlation between aspects of drawings are described. The finding of trends in correlation is the main result of this paper.
Polanyi's theory of consciousness [12,13], relates small scale and large scale elements of experience to each other. This relation is generalised (with a newly defined concept called 'breath' in paragraph 3) and applied to describe the differences between the trends. This is possible because the aspects defined for the drawings allow for grouping in small and large scale aspects with regard to the scale of the drawing itself. Also the breath concept is valuable as an intermediary between the intentions and attitudes of the artist, and the measured aspects of the drawing.

Each trend in correlation can be summarized in a specific diagram, and names can be attached to those of them that allow for the introduction of analogies. In the text the words house and splinter are chosen that have a specific spatial and temporal meaning with both a modern and a historical component. In this way the meaning of the existing trends in correlation and the related trends in concepts of space and time can be made more precisely and open to discussion.

The outline of this paper is as follows. In paragraph 2 possible aspects of drawings are considered and defined. A series of aspects is chosen for further study because they are related to the motor experience of the artists. Correlation of the aspects as it occurs for the investigated drawings is explained in relation with a generalisation of Polanyi's theory of consciousness in paragraph 3. In order to do this the concept of breath is introduced. In the discussion section paragraph 4 artist's attitudes are linked with drawing aspects and motor experiences through the concept of breath. For the more complex aspects, the used models they are derived with are explained in appendix A. The counting of curved bends and corner bends is defined in appendix B.

2. ASPECTS OF DRAWINGS AND THE MOTOR RECOGNITION SUBGROUP

For this study drawings are chosen that are made by physics students for practicum reports for the first and second year undergraduate curriculum at Utrecht University in the years 2002/2003. These drawings are expected to be exemplary for drawings made by physics students and physicists in laboratory diaries during experiments. Also many of them are available for a series of experiments and thus they are a comparable set. What is specific about these drawings is that they represent the experimental situation as it is created by the experimenter. The experimental set-up is designed to produce data that can be placed in a theoretical context.

Investigated are aspects that can be considered to be the building blocks of a drawing and that can be counted or measured. This is done in order to be able to compare aspects, and search for possible correlation. Elementary aspects like building blocks can inform one about how the one who made the drawing experiences elementary notions like plasticity, spatiality etc., in the context of the experimental situation and its theory.

A summary of several aspects that are interesting for this research is listed below. What is concentrated on is the fact that all drawings are line drawings. Some of the aspects are descriptive in character, but most of them are aspects that can be given a value.

1. Colour or black and white, 2. two- or three-dimensional rendering, 3. use of perspective, 4. maximum size (large scale size), 5. minimum size (small scale size), 6. surface character of the drawn objects, 7. interior character of the drawn objects, 8. line-diversity including presence of plasticity or spatial elements, 9. total number of ends and bends, where bends are considered to be a special form of ends, for all line-segments, 10. percentage of soft bends, 11. percentage of corner bends, 12. number of independent forms, 13. size of main independent forms, 14. number of background forms, 15. composition resembling portrait, still-life, landscape or other, 16. distribution of independent forms, for instance linear, 17. Content, that is experimental set-up content, theoretical content or both, 18. real size of the experimental set-up selected to be drawn.

The above aspects are all general aspects descriptive of line drawings. Here, concentrated on are recent original line drawings made by the experimenters, as objects of study. Therefore among the objects not chosen are photographs of the experimental set-up situation, standard textbook drawings and visualization of measurement results. Some of the aspects, that are not chosen for this investigation but which might be interesting in their own right, are: metaphorical/ pictorial content and artists aspects like social or cultural background or gender.

Besides the listed aspects, also there can be defined several complex aspects that are derived from the above aspects as functions of the values of some of them. These complex aspects are defined in such a way that they are measures for the expressiveness of a drawing, due to different models: when looking at line drawings, one can define in different ways the flow of attention for how one follows the lines that make up the drawing. The models are explained in appendix A. The following three aspects can be introduced: aspect (a) as the microscopic local density of the configuration of line-segments in the drawing, and corresponding to it a time scale, local displacement time, as the time relevant for displacement over a standard local distance, defined by a diffusion model. Aspect (b) as the time necessary for observing the drawing as one whole, following a series of lines throughout the drawing, defined by a kinematic model. Since aspect (a) and (b), that are both time related, are small and large scale aspects respectively, they are understood to be independent of each other. Aspect (c), nearness, that is, how near to you an object in a drawing literally seems to be, is defined by the size of the main independent form and line-diversity and the inverse of the minimum size, the last two aspects indicative for detail. Aspects (a) and (b) are 'in the plane' properties of the drawing. Aspect (c) is a 'perpendicular to the plane' element aspect. Both extremely high or low values for aspect (a), (b) or (c) mean 'outspoken character' and are taken as an indication of expressiveness.

From the listed aspects a subgroup is chosen that consists of the aspects that are related to motor recognition. This is done to investigate the influence of the fact that these drawings are made by the same experimenter who constructed the experimental set-up. Reaching, lifting and touching are basic physical actions that each have a specific connection to how physical space and time are perceived. It is assumed that these perceptions are related to both the experimental set-up design and the theoretical context. In relation to this, the building of the experimental set-up in reality from its composing parts is a decisive motor experience itself that will influence the way the experimental situation will be represented in a drawing.
To this subgroup are added the complex aspects, that all three are derived from the motor recognition aspects, and the aspects for content and real size. The aspects (a) and (b) are related to the time behaviour during drawing, of the one who made the drawing. With time behaviour is meant how in time the line-segments and line-segment configurations in a drawing are made. On the level of line segments in a drawing this means that aspects (a) and (b) describe how strokes are made and perceived in time. Aspect (c) is related to 'nearness' behaviour, it describes how strokes are made in space and on the near, or remote, drawing surface. These attitude aspects are exemplary for the influence of behaviour in general, or attitude, on drawing as it occurs in the other aspects as well.
The subgroup of aspects then consists of the following. Values related to reaching: the large and small scale size aspects. Values related to weight and to lifting: line-diversity and together with it plasticity and spatiality elements present in the drawing, and surface and interior characteristics. Values related to collisions and touching: the total number of ends and bends and the percentage soft bends versus corner bends, in relation with how ends and bends, generally speaking discontinuities, are made, discontinuities in motion being related to collisions and touching. The counting of numbers of bends is defined in appendix B. Aspects for expressiveness: (a) local displacement time, small scale line particulars, (b) recurrence time, large scale line particulars and aspect (c) nearness. Finally, the aspect for content (experimental set-up content, theoretical content or both) and the real size aspect for the real experimental set-up as far as it is drawn.

3. TRENDS IN CORRELATION AND POLANYI'S THEORY OF CONSCIOUSNESS GENERALISED

The main result of this study is that trends of correlation between the aspects of the above subgroup are found to exist. One particular kind of correlation trend is named the group of natural correlation. It is the main trend among the investigated drawings. As an example the following two clusters are given that represent two sets of aspect values that correlate positively within this group. Every drawing that belongs to the natural correlation trend can be attributed to one of these clusters. Cluster I: small total number of ends and bends, theoretical content, spatially inclined, large percentage of soft bends, low aspect (a) local displacement time, low aspect (b) recurrence time. Cluster II: large total number of ends and bends, experimental set-up content, plastically inclined, large percentage of corner bends, high aspect (a) local displacement time, high aspect (b) recurrence time.

To facilitate generalization, a correlation group can be represented in a diagram that involves all aspects of the subgroup. The natural correlation group that is considered in the first diagram (diagrams are included at the end of the paper) is only one group of the possible correlation groups between the ten aspects of the subgroup. In the diagram an off-diagonal + does not mean high value, only positive correlation with other +'s and negative correlation with off-diagonal –'s. The aspects are defined in such a way that for the natural correlation diagram mostly +'s occur. These correlations are natural in the sense that they are expected and easily imagined to be true. The only minus occurs for the small scale size correlation with, for instance, the large scale size. Larger drawings seem to favour smaller details. This correlation can be explained by considering the notion of dwelling.

In a larger drawing space the artist feels more free and safe from the boundary to concentrate on finer details. Dwelling as it can be defined specifically for the natural correlation diagram with high large scale aspect (b), recurrence time, (cluster II) is giving your attention to a drawing for a longer time while retreating freely from detail to smaller detail and covering a large space, and time, in this manner.
This property of the drawing, or equivalently this activity by the person who made the drawing and also of someone looking at the drawing, provides a reason for why the above mentioned correlation between large scale size and small scale size in the natural correlation diagram can occur. Dwelling as a means of bridging the gap between the large scale size and the small scale size resembles in-dwelling discussed by Polanyi as describing consciousness. In-dwelling is the same as going from a focus on the details to the large scale form, relying on the details like one relies on ones consciousness of our bodily movements for achieving a coherent action of the body as a whole. The small scale interpretation is relied upon while the focus is upon the large scale interpretation.

Every different diagram represents a different correlation group. For every correlation group there can be drawings to be placed in the whole range from extremely high to extremely low values for every aspect as long as the correlation between the aspects given by the diagram is adhered to. Every specific diagram can be assigned a word, a concept, that is related to one specific choice of high or low values for the aspects. As two examples diagrams related to the words splinter and house are explored. The diagonals represent the properties for the words splinter and house with (+) meaning high value, (-) low value. These example diagrams are incomplete. Finding more properties for these words makes the diagrams more complete but one has to be careful not to restrict the words in their meaning. At least the natural correlation diagram describes a set of real drawings. In fact most drawings of the ones investigated belong to this set. Many other, maybe incomplete, diagrams are imaginable. Natural correlation corresponds to the general trend, however drawings that deviate from it exist in the investigated collection.
The diagram for the word splinter with the properties high plasticity (in the diagram designated with (+) for plasticity/spatiality), high aspect (b) recurrence time and low small scale size, resembles the natural correlation diagram (cluster II). However when the property low large scale size is added it deviates from it: the natural correlation group itself includes the correlation of higher large scale size with lower small scale size instead of lower large scale size with lower scale small size.
Indeed some drawings exist among the investigated ones, that have the properties attributed to the splinter diagram. These drawings, thus deviating from the ones that have the natural correlation trend, can be described by the fact that dwelling is difficult to occur when looking at them.
Because splinter diagram drawings have lower large scale size but also high aspect (b) recurrence time they are very concentrated and restrained. The diagram for the word house, as far as it is defined within the diagram, with only the properties high spatiality and low aspect (a) local displacement time does not deviate from the natural correlation diagram.

Not all results can be included in the diagram. This is true for the result that drawings can be divided into two groups, one with the absolute number of soft bends smaller than 12 and one with this number larger than 12. The first group has a relation between the total number of ends and bends and the percentage of soft bends that differs from the one of the second group. This is a kind of result that cannot be expressed by these diagrams.

The correlation group diagram differences make clear that for each diagram a specific concept of attention is needed. For instance for the natural correlation group the above mentioned dwelling property that higher large scale size correlates with lower small scale size suggests a sphere concept of attention. This includes a constant total volume of attention, say a 'breath', dispersed over spheres with sizes / volumes related to the small scale size in an outer volume with a size related to the large scale size. For a larger outer volume the spheres of attention will increase in number but decrease in size keeping the concentration of spheres in the volume constant.
This concept of attention has to be changed when considering the splinter diagram drawings. There the concentration of spheres of attention will have to increase when decreasing both the outer volume and the sphere size or the constraint of constant total sphere volume, constant breath, has to be abandoned.
One may propose that other aspects can be coupled to the elements that construct the concepts of attention. For instance one may relate increasing sphere volume or decreasing concentration to increasing theory content when theory is considered to unite and cross distances, as opposed to the experimental set-up considered to demand mechanical contact. This seems reasonable for the natural correlation diagram drawings. The two drawings in the investigated collection that correspond to the splinter diagram are of experimental set-up content. Both have a different way of dealing with distances, one seems to be a miniature, shrinking to smaller sizes, the other is as if it is curled up to a small size. For both drawings it appears as if they are only a leftover part or detail of a larger drawing, a fragment that in its content repeats or recapitulates (nearly) the original that once existed before as a whole, a fragment that because of its detail asks for time, and care, to look at it. This is the reason for choosing the word splinter for these diagram drawings [14].

The house correlation group has emphasis on correlation of high spatiality and low aspect (a) short displacement time. For the natural correlation group this correlation is extended towards high theoretical content. This would mean within the breath concept above that house diagram drawings have increasing sphere volume together with decreasing concentration. However house diagram drawings properties cannot be generalized in this way. For instance, it is feasible that one finds also high plasticity in these drawings. Then also experimental set-up and a decreasing sphere volume would result, giving a combination of spheres with varying volumes and making the breath concept very complex.
It depends on the existing drawings for which the word-concept house seems appropriate, which properties are defined with the word. The historical meaning of the word house, for instance as abode or dwelling-place [14], gives opportunities to develop further the concepts of attention and breath attached to it. Restricting oneself to the drawings that were investigated one finds that indeed both drawings with only a spatiality component and drawings with a spatiality and a plasticity component correspond well to the meaning of the word house as dwelling-place. However plasticity seems needed to give to a house drawing a stillness and peace that is lacking from the drawings with only spatiality. These latter drawings resemble mostly a dwelling-place in open air without roof.
In the diagram, adding plasticity would result in changing the (-) into (±) for plasticity/spatiality and changing correspondingly the off-diagonal parts. One notices that for an original dwelling-place stillness, resting, is combined with motion, living. This suggests that together with plasticity also high large scale aspect (b) recurrence time should be included in the house diagram. The word-concept house as abode or dwelling-place clearly relates in this way to in-dwelling: the relatively large scale motion aspect of the drawing combines with the relatively small scale stillness aspect. However the small scale stillness aspect is not experienced unconsciously. Attention seems to be divided over small scale and large scale aspects and the movement of attention is thus more complex than for the drawings that do correspond to in-dwelling proper.

4. DISCUSSION: THE LINK BETWEEN MOTOR RECOGNITION AND MEANING

Attitudes can be introduced to describe the behaviour of someone who makes the drawing. Because aspects (a) and (b) are related to time they make conceivable the time attitude during drawing when producing strokes of line-segments and line-segment configurations. Low aspect (a) local displacement time, because different parts in the drawing are easily reached, is related to a unity attitude, that gives value to a feeling of completeness and immediate unity because a final and decisive result is derived at. High aspect (b) recurrence time is related to a time and care attitude to describe that a longer time and more caring is needed to complete or observe the drawing as a whole.
Because both aspect (a) and (b) are time like aspects they are both related to two, specific and differing, kinds of viewing times. Viewing times are a measure for how long a drawing catches one's attention. Very short or very long viewing times are an indication for high expressiveness. Thus extreme variants of the two time attitudes (unity attitude, and care and time attitude) are related to expressiveness of the drawing.

The meaning and intention of the one who makes the drawing are demonstrated, 'performed', by the attitudes that are displayed in the drawing. This is an important relation between both the experimental set-up design and theoretical context of the experiment, and the aspects of the drawing. Central to this relation is the consciousness-unconsciousness link 'breath': breath is the link between known and un-known, between unconscious motor perception and conscious perception and thinking and aiming. Breath is, in this way, generating meaning into drawing as an action, as a performance [ref.15]. Also it is clear now that breath is the key element when translating attention from the performer (the artist who is drawing) towards the observer (someone who is viewing the drawing). For both of them the breath will be the same. That is why in this article no difference is made between the flow of attention for both of them. Movement of attention between large scale elements of the drawing and small scale elements can be described by a generalisation of the concept of in-dwelling in the theory by Polanyi. This description depends on whether a drawing belongs to the natural correlation trend group or to another group. Every correlation trend group has its own specific definition of its large scale to small scale movement of attention. The natural correlation trend group is specified by the constraint of constant breath: for increasing large scale sizes (large form measure) there are decreasing small scale sizes (detail measure) as explained in paragraph 3.

Above, attitudes were attached to low aspect (a) or high aspect (b) respectively. Since these aspects are time-like they are suited very well for relating to attitudes directly, as the time behaviour and time experience are basic notions for these. Attitudes can be enriched by considering all other aspects of the considered subgroup of motor recognition aspects as well. Through these aspects one is concerned with presence (plasticity, corner bends, high aspect (c)), distance (the size aspects, spatiality, soft bends), weight and contact (plasticity, ends and bends in general) or content, and thus the attitudes also acquire these characteristics. In the case of natural correlation, according to the place within the two clusters of this correlation group, there follows a distribution of elements of the attitudes in relation to a distribution of aspects:

| Cluster I | Attitude elements | Cluster II | Attitude elements |
| --- | --- | --- | --- |
| Low large scale size | Small distance | High large scale size | Large distance |
| High small scale size | Large detail distance | Low small scale size | Small detail distance |
| Spatiality | Distance | Plasticity | Weight, contact, presence |
| Low number of ends and bends | Distance | High number of ends and bends | Weight, contact |
| High % soft bends | Distance | High % corner bends | Presence |
| Low aspect (a) | **Unity** | High aspect (a) | Scattered-ness |
| Low aspect (b) | Directness | High aspect (b) | **Time and care** |
| Low aspect (c) | Remoteness | High aspect (c) | Presence, nearness |

When considering now for instance the house diagram with aspects high spatiality and low aspect (a) one finds from the above schema the attitude characteristics distance and unity. Adding the aspects plasticity and high aspect (b) would result in leaving the cluster I / II division in the schema and adding attitude characteristics presence, contact, weight and time and care.
The splinter diagram with aspects low large scale size, low small scale size, plasticity and high aspect (b) can be related to the mixed attitude characteristics as they are described above. In this way very complex attitudes can be characterised.
Different correlation groups will have different characteristic combinations of large scale and small scale aspects for drawings belonging to these groups. Also the attitude characteristics as summarized above are distributed in a proper way to agree with the correlation group. This determines the different characteristic movements of attention related to these correlation groups and attitudes. The sphere model of consciousness and the concept of constant breath as it is introduced in paragraph 3 is only one of many possibilities that relate to the correlation groups. The correlation group attitudes, and the word analogy and the concept of breath attached to it, thus describe the experimental set-up design and theoretical context of an experiment in a meaningful way.

This study was made possible with support of The Netherlands Foundation for Visual Arts, Design and Architecture.

Natural correlation group:

| | large size | small size | plasticity /spatiality | ends and bends | corner/soft bends | set-up /theory | real size | aspect (a) | aspect (b) | aspect (c) |
|---|---|---|---|---|---|---|---|---|---|---|
| large size | x | - | + | + | + | + | + | + | + | + |
| small size | - | x | - | - | - | - | - | - | - | - |
| plasticity / spatiality | + | - | x | + | + | + | + | + | + | + |
| ends and bends | + | - | + | x | + | + | + | + | + | + |
| corners / soft bends | + | - | + | + | x | + | + | + | + | + |
| set-up / theory | + | - | + | + | + | x | + | + | + | + |
| real size | + | - | + | + | + | + | x | + | + | + |
| aspect (a) | + | - | + | + | + | + | + | x | + | + |
| aspect (b) | + | - | + | + | + | + | + | + | x | + |
| aspect (c) | + | - | + | + | + | + | + | + | + | x |

Splinter properties: (+) high plasticity, (-) low large size, (-) low small size and (+) high aspect (b) recurrence time

|  | large size | small size | plasticity /spatiality | ends and bends | corner/soft bends | set-up /theory | real size | aspect (a) | aspect (b) | aspect (c) |
|---|---|---|---|---|---|---|---|---|---|---|
| large size | (-) | + | - |  |  |  |  |  | - |  |
| small size | + | (-) | - |  |  |  |  |  | - |  |
| plasticity / spatiality | - | - | (+) |  |  |  |  |  | + |  |
| ends and bends |  |  |  | x |  |  |  |  |  |  |
| corner / soft bends |  |  |  |  | x |  |  |  |  |  |
| set-up / theory |  |  |  |  |  | x |  |  |  |  |
| real size |  |  |  |  |  |  | x |  |  |  |
| aspect (a) |  |  |  |  |  |  |  | x |  |  |
| aspect (b) | - | - | + |  |  |  |  |  | (+) |  |
| aspect (c) |  |  |  |  |  |  |  |  |  | x |

House properties: (-) high spatiality, (-) low aspect (a) local displacement time

|  | large size | small size | plasticity /spatiality | ends and bends | corner/soft bends | set-up /theory | real size | aspect (a) | aspect (b) | aspect (c) |
|---|---|---|---|---|---|---|---|---|---|---|
| large size | x |  |  |  |  |  |  |  |  |  |
| small size |  | x |  |  |  |  |  |  |  |  |
| plasticity / spatiality |  |  | (-) |  |  |  |  | + |  |  |
| ends and bends |  |  |  | x |  |  |  |  |  |  |
| corners / soft bends |  |  |  |  | x |  |  |  |  |  |
| set-up / theory |  |  |  |  |  | x |  |  |  |  |
| real size |  |  |  |  |  |  | x |  |  |  |
| aspect (a) |  |  | + |  |  |  |  | (-) |  |  |
| aspect (b) |  |  |  |  |  |  |  |  | x |  |
| aspect (c) |  |  |  |  |  |  |  |  |  | x |

APPENDIX A: DIFFUSION MODEL AND KINEMATIC MODEL FOR LINE CONFIGURATION DRAWINGS

In this appendix the diffusion model and the kinematic model that are used to derive the values for the aspects (a) and (b) respectively will be explained. All drawings investigated can be described as line drawings, that is as line configurations, with line segments that can vary in thickness and shape or wrinkling. Both the diffusion model as the kinematic model describe observing, or creating, a line drawing. Line segments are focused upon one by one in a consecutive order. Experiencing the line segments is identified within these models with the stream of line segments that arises from diffusion of attention that is directed towards the line segments, or from the kinematics of the stroke of the line segments. Whereas the kinematics of a stroke is concrete and real, and a series of strokes can be directly related to a time interval, with the diffusion model it is the experience, because of looking at the line segments, and the attention that moves along the line segment configuration, that is valued indirectly. In the diffusion model properties of diffusion like temperature and density are related to a time interval, local displacement time, for line segments locally. In this way for both models time related values are derived, that are used to quantify aspects (a) and (b).
1) The quality (a) relates to time intervals that describe the composition of line-segments that form a more or less dense, criss-cross like configuration locally on the scale of the line-segment. Quality (b), recurrence time, relates to the time t it takes to follow a series of line-segments that form a melody throughout the drawing. Nearness quality (c) relates to simply how near a drawing is experienced to be. This last is quantified by taking the product of the main independent form size, line diversity and the inverse of the minimum size. When regarding a drawing as diffusing the attention of the observer of the drawing along some line-segments through a configuration of these, elements resembling density, temperature and time can be attributed to it. It has to be remembered that these elements belong to the diffusion model, and have to be translated before being used to describe drawings.
For diffusion the following equations are well known. The mobility equation $v = B K$ relates the velocity v with the inverse shear B and exerted force K. The diffusion equation $j = -D\, dn/dx$ relates the stream j with the change in the density $dn/dx$ (D diffusion constant). Then from the Einstein-law $D = B\, kT$, with kT the temperature, and using time $t = R / v$ with R the covered distance it follows that $t = R\, kT / D\, K$.
Now a simple identification follows from the interaction between the one viewing the drawing and the drawing itself. The stream j can be related to the stream of experiences because of looking at the individual line-segments in the drawing and the density n to the intensity of experiencing a line-segment or part of the drawing. In this way one arrives at the following definitions: j = (length of line-segment) / (time for experiencing the line-segment) or abbreviated j = (l-segment) / $t_0$, and $dn/dx$ = -(line-diversity) / (l-segment). An equation for $t_0$ in terms of aspects of a drawing is derived in part 2) of the appendix. It depends on the percentages of soft bends and corners in the drawing. In this way D can be expressed in terms of aspects of the drawing.
K is taken to be constant (the exerted force relates to the observer) and R is taken to be equal to the total number of line-segments (#) times the length of a line-segment. Combination of the equations above then gives the result $t = \text{constant}\, kT\, t_0\, \#\, (\text{l-diversity}) / (\text{l-segment})$.
The total time should be equal to $t_0$ times the number of line-segments (#), thus is approximated
$t = t_0\, \#$ (for convenience the number of ends and bends is taken to be the same as the number of line segments). At first sight this value for time t is not exactly the same as the time defined above as aspect of expressiveness (b) because there is only one large scale series of line segments that is singled out from the total of all line segments and one would expect the one-dimensional value $t = t_0\, \#^{1/2}$. However following only the series one still experiences the total drawing with all its line-segments, because due to its special expressive character, that singled out series has a range that covers the total drawing with all its line-segments. The expressive line-segment is sensitive to its surroundings. This causes the series to be a real melody through the drawing. Thus $t = t_0\, \#$ is a fairly good approximation since the large scale property of this aspect of expressiveness is represented by it.
For the temperature kT the resulting equation is $kT = \text{constant}\, (\text{l-segment}) / (\text{l-diversity})$. Notice that kT is proportional to the inverse density change $1/(dn/dx)$. The density n itself can be integrated for constant average line-diversity and line-segment length to give $n = L\, dn/dx$ for a trajectory length comparable to the large scale size (L) of the drawing. Thus n is a large scale (L) density while kT is an inverse microscopic density. This is as it should be because temperature is usually related with inverse density through the kinetic gas-law and is itself microscopic in character.
Translation to the aspects of drawings can then be made as follows. Quality (a) relates to the microscopic density proportional to 1/kT for line-segment configurations and is described by local displacement time scales that correspond to the time relevant for displacement over a specific distance that will be more for higher density. Low local density describes a unified and connected drawing with short time intervals while high local density describes disconnection and slowing down of line segment and drawing experience.
Quality (b), the recurrence time, is described by the time t=t0 # where short time t relates to directness in experiencing a drawing and long time t relates to taking much more time and thus giving much more care to a drawing.

2) Time t0 per line-segment with a soft bend. Now it remains to estimate the time t0 for line-segments that contain or border on soft bends or corner bends. For line-segments with soft bends literature exists and this can be extended to line-segments with corner bends.
For the description of a line segment there was defined t0 as the time that passed during drawing, during the stroke of the line segment. t0 is taken not between two successive bends but however around one bend so t0 = t(x+1)-t(x-1) where x denotes the position of the bend and +/-1 half of the stroke. One bend is one 1/4 th part of a circle (soft bend, or a rectangle :corner bend) with radius R around O and with x0 at the centre of the bend from x-1 t x+1.
For soft bends the theory in the literature [Adi-Japha,16] is followed. There the relation
$A(t) = K_1 C(t)^{2/3}$ is derived with A(t) angular velocity and $C(t) = |v_x a_y - a_x v_y| / (v_x^2 + v_y^2)^{3/2}$ the curvature of the stroke drawn at time t with velocity v and acceleration a and K1 a constant.
For every part of the circular (soft bend) stroke with radius R and co-ordinates (x, y) there is C = constant = $1 / (x^2 + y^2)^{1/2} = 1/R$ and the angular velocity is $A = K_1 C^{2/3} = K_1 (1/R)^{2/3}$ = real velocity/R. Thus we have for the stroke of length R the time t0 = (stroke-length) (1/real velocity) = $K_1 R^{2/3}$ where all numerical constants are included in a new constant K1.

3) Time t0 for corner bends. The above equation for the stroke time for soft bends does not apply for corner bends. The time t0 is defined as an addition of the time along the sides before and after the corner itself and the time t0(x0) during the corner: t0 = t0(side1) + t0(x0) + t0(side2). If one takes for t0(x0) the time for a soft bend with a limit R = 0 it follows that t0(x0) = 0.
During the stroke both the velocities and accelerations are assumed to be approaching zero towards the corner itself, that is the stroke slows down before changing direction (side1) and starts with slow acceleration in the new direction (side2). The direction change at (x0) is taken to be a discontinuity caused by for instance the presence of a second object in contact with the object the stroke was applied for in the drawing. Also assumed is that for side1 acceleration a is proportional to minus velocity v and for side2 acceleration a is proportional to velocity v. For convenience all components are taken to be the same and equal to +/-r. For a corner bend with side length R now is defined the angular velocity = $(v_x^2 + v_y^2)^{1/2}$ 1/R = $(r^2 + r^2)^{1/2}$ 1/R.
The curvature C is different for soft bends because there v and a are perpendicular and for corner bends they are parallel. Therefore an alternative curvature C' is defined $C' = |v_x a_x - a_y v_y| / (v_x^2 + v_y^2)^{3/2}$. C' is defined in such a way that its value is the same as for a soft bend with a radius R when equating the angular velocity A again to $K_2 C'^{2/3}$, with K2 a new constant. It then follows that the real velocity has the value $(r^2 + r^2)^{1/2} = (K_2 R)^{3/5}$ or $r = (1/2)^{1/2} (K_2 R)^{3/5}$.
The stroke time for side1 or side 2 is taken to be t0(side1 or 2) = R 1/real velocity = $(1/K_2)^{3/5} R^{2/5}$. Including all numerical constants again in K2 the result is t0 = t(side1) + t(side2) = $K_2 R^{2/5}$. For R can be inserted the average line-segment length (l-segment). Taking together all soft bends and corner bends for one drawing results in the average time interval t0:
t0 = (%soft bends)$K_1$(l-segment)$^{2/3}$ + (%corner bends)$K_2$(l-segment)$^{2/5}$

4) Denoted with (l-segment) is the average line-segment length for one drawing and this is approximated by the value (l-segment) = $( L * s )^{(1/2)}$, (L) and (s) being the large and small form size respectively. For some specific groups of drawings (l-segment) = $( L * s )^{(1/2)}$ tends to be, in a very approximate trend, constant and independent of (L) or the total number of bends and ends.

5) To quantify the average line-diversity for a drawing, where lines are diverse with respect to thickness or made up of multiple parts or scribbles, it is combined with the presence of spatial or plasticity properties (a line is drawn as being a border to an open space one can look into, respectively a line that is drawn as a border to a closed space one cannot look into or behind).
Now to each drawing a number is attributed: 2 when none of these properties are present and division by 2 for spatiality and multiplication by 2 for plasticity and diversity (so spatiality combined with plasticity and diversity in one drawing gives the number 4).

The reason for this combined definition is that spatiality is directed outwards from one line-segment mostly to other line-segments whereas plasticity and diversity are experienced from within as contributing more weight to a line. Besides this there is multiplied with the percentage of real ends. Real ends are related to the occurrence of eclipses of one form by another form and therefore they are a measure for presence of plasticity presence. This is only a very simple way of attribution. All kinds of differences between different parts of one drawing are not regarded. However for average properties like time (t), for a drawing seen as one unity, it seems justified.

APPENDIX B: COUNTING BENDS IN A CONFIGURATION OF LINES

There is not just one unambiguous way to count bends in a configuration of lines. From the drawing alone one cannot infer where one line ends and another line begins. Is there continuation of stroke from one line-segment to another line-segment or not? Sometimes it is unclear whether one is looking at a combination of ends and corners or ends alone or corners alone. Still, below is made an attempt to define a counting rule for junctions of line-segments. This is not the only possible counting rule. However it is one that combines well with the kinematic model introduced in appendix A and also is intuitionally appealing.

Corner bends are the bends that represent an abrupt change of direction. Within the kinematic model this is interpreted to mean that, when drawing these corner bends, one slows down in the original direction and then returns to the original momentum of drawing in the new direction. This is different for soft bends where there is no slowing down when drawing the curve the soft bend occurs in.
To count corner bends one might define a junction where three or more line-segments meet from three or more different directions that are each other's continuation, with the exception of maybe one ending line-segment. This leads to the following counting convention for corner bends together with real ends.
Definition: a junction of line-segments counts equal to the number of line-segments that participate minus one and minus one for every line-segments that is the straight continuation of another line-segment, because it has a corner of 180 degrees with it and minus one for a line-segment that has a real end at the junction because it is without any continuation at all.
When two line-segments meet at a junction and have as straight continuations at 180 degrees two other line-segments this is counted as one corner bend even though the lines themselves make no corner bends but only cross each other. The argument for this is that these crossing line-segments, when they are drawn, are assumed to influence each other by slowing down each other just like for real corner bends.
Soft bends are the curves of line-segments. Definition: soft bends are counted by taking four curves to form one circle. Thus one soft bend curve should approximately cover 90 degrees. This counting of curves is independent from whether there is a contact with another curved line-segment or with a straight line-segment.

REFERENCES

[1]  Tufte E., The visual display of quantitative information, Graphics Press, Cheshire Connecticut 1984

[2]  Woolgar M. Lynch S., (editors), Representation in scientific practice, MIT Press 1990

[3]  Baigrie B., (editor), Picturing knowledge, University of Toronto Press 1996

[4]  McLean M., Henson Q., Hiles L., The possible contribution of student drawings to evaluation in a new problem-based learning medical programme: a pilot study, Medical Education 37 2003 (895)

[5]  Gordon D., Edelson D., Pea R., symposium scientific visualization tools in science classrooms, annual meeting of the American Educational Research Association, New York 1996

[6]  Harrison A., Treagust D., Learning about atoms, molecules and chemical bonds; a case study of multiple-model use in grade 11 chemistry, Sci. Ed. 84 2000 (352)

[7]  Pittman K., Student-generated analogies, another way of knowing, J. Res. Sci. Teach. 36 1999 (1)

[8] Read H., Education through art, Pantheon Books, New York 1958

[9] Hesse M., Models and analogies in science, Newman history and philosophy of science series 14, Sheed and Ward, London, 1963

[10] Hesse M., The structure of scientific inference, Macmillan Press, London 1974

[11] Kojevnikov A., Freedom, collectivism, and quasiparticles: Social metaphors in quantum physics, Historical studies in the physical and biological sciences, 29 (part 2) 1999 (295)

[12] Polanyi M., Personal Knowledge, Routledge & Kegan Paul, London, 1958

[13] Innis R,. Consciousness and the play of signs, Indiana UP, Bloomington and Indianapolis 1994

[14] Oxford English Dictionary, entry splinter XVI p. 285, entry house VII p. 435, J.A. Simpson, E.S.C. Weiner, second edition, Clarendon Press, Oxford, 1989

[15] Nair S., Restoration of Breath: Consciousness and Performance, Rodopi, Amsterdam/New York, 2007

[16] Adi-Japha E., Levin I., Solomon S., Emergence of representation in drawing: the relation between kinematic and referential aspects, Cognitive Development 13 (1) 1998 (25)